# Monitoring breast cancer response to neoadjuvant chemotherapy with ultrasound signal statistics and integrated backscatter

Hanna Piotrzkowska-Wróblewska[1], Katarzyna Dobruch-Sobczak[1, 2], Ziemowit Klimoda[1], Piotr Karwat[1], Katarzyna Roszkowska-Purska[3], Magdalena Gumowska[2], Jerzy Litniewski[1*]

[1]Ultrasound Department, Institute of Fundamental Technological Research, Polish Academy of Sciences, Warsaw, Poland

[2]Radiology Department, Cancer Center and Institute of Oncology, M. Skłodowska-Curie Memorial, Warsaw, Poland

[3] Department of Pathology, Cancer Center and Institute of Oncology, M. Skłodowska-Curie Memorial, Warsaw, Poland

*jlitn@ippt.pan.pl (JL)

# Abstract

Neoadjuvant chemotherapy (NAC) is used in patients with breast cancer to reduce tumor focus, metastatic risk, and patient mortality. Monitoring NAC effects is necessary to capture resistant patients and stop or change treatment. The existing methods for evaluating NAC results have some limitations. The aim of this study was to assess the tumor response at an early stage, after the first doses of the NAC, based on the variability of the backscattered ultrasound energy, and backscatter statistics. The backscatter statistics has not previously been used to monitor NAC effects.

The B-mode ultrasound images and raw radio frequency data from breast tumors were obtained using an ultrasound scanner before chemotherapy and 1 week after each NAC cycle. Twenty-four malignant breast cancers, qualified for neoadjuvant treatment before surgery, were included in the study. The shape parameter of the homodyned K distribution and integrated backscatter, along with the tumor size in the longest dimension, were determined based on ultrasound data and used as markers for NAC response. Cancer tumors were assigned to responding and non-responding groups, according to histopathological evaluation, which was a reference in assessing the utility of markers. Statistical analysis was performed to rate the ability of markers to predict the final NAC response based on data obtained after subsequent therapeutic doses.

Statistically significant differences between groups were obtained after 2, 3, 4, and 5 doses of NAC for quantitative ultrasound markers and after 5 doses for the assessment based on maximum tumor dimension. After the second and third NAC courses the marker, which was a linear combination of both quantitative ultrasound parameters, was characterized by an AUC of 0.82 and 0.91, respectively.

The introduction of statistical parameters of ultrasonic backscatter to monitor the effects of chemotherapy can increase the effectiveness of monitoring and contribute to a better personalization of NAC therapy.

# Introduction

Neoadjuvant chemotherapy (NAC) was initially used in locally advanced breast cancer (LABC) and in the case of inflammatory cancer [1]. Currently, it is also recommended in patients with triple-negative cancer (TNBC), with the presence of HER-2 + receptors (Luminal B HER2-positive and HER-positive non-luminal subtype), and in cases of luminal B HER2-negative tumors with low expression of hormone receptors, with high grade of malignancy (G3) in patients at an early age (≤ 35 years) in the second or third stage of cancer [2–4]. Neoadjuvant chemotherapy reduces the risk of metastases and micrometastases in distant organs. It also reduces the neoplastic focus and decreases the frequency of the recurrences and the mortality of patients.

The use of NAC therapy does not always bring the expected results. A meta-analysis conducted on a group of over 18,000 patients (from 49 studies) who received NAC showed that the pathologic complete response rate was 21.5% [5]. In the HER2-positive disease and the triple-negative disease groups, the percentages of complete responses for the treatment reached 76 and 67%, respectively [5]. The complete response is significantly lower in the case of the luminal B HER2–negative subtype, but patients with additional risk factors may also benefit from pre-operative treatment. According to available data, even over 40% of patients undergoing chemotherapy show a poor pathological response to the treatment [6]. In such cases, the whole cycle of neoadjuvant therapy, which lasts about 4 months (five cycles), is associated with the delayed start of another, more effective treatment and unnecessary exposure of the patient to the toxic effects of the drugs used. The response to the treatment varies and requires differentiation between responders and non-responders.

For monitoring, a clinical breast examination (CBE), mammography (MMG), traditional ultrasound imaging in B-mode (US), and magnetic resonance imaging (MRI) are used. Imaging with MRI is more accurate compared to CBE, US, or MMG; however, MRI is a technique with limited availability. Ultrasonography is considered a more accurate method in assessing tumor size and in the monitoring of residual breast tumors than CBE or MMG [7]. Recent work has also shown that classical US imaging techniques with sonoelastography are useful and allow predicting the response to the treatment with a high degree of sensitivity and specificity after the second course of chemotherapy. The decrease in tumor stiffness is a good predictor of a pathological response [8].

Methods based on monitoring changes in tumor size, which are determined on the basis of imaging tests or by palpation examination, have numerous limitations [7]. Changes in tumor size occur with a delay in relation to changes in the tumor microstructure. Sometimes, despite a positive pathological response to the treatment, there is no apparent reduction in primary tumor mass in imaging because it is masked by changes induced by the treatment. Functional imaging techniques, such as positron emission tomography, MRI with contrast agents, diffusion weighted imaging, and diffuse optical spectroscopy, enable to capture changes in the microstructure, vascularization, and metabolic activity of tumors under the influence of chemotherapy after the first cycle of treatment [9–11]. However, these are cost-intensive and time-consuming methods which require intravenous administration of exogenous contrast agents to detect changes in the tumor after each course of chemotherapy. Therefore, their use is limited.

The use of quantitative ultrasound (QUS) methods to assess the effectiveness of chemotherapy seems to be an interesting alternative to tracking changes in tumor size. The QUS techniques are based on the analysis of raw, ultrasonic radio-frequency echoes (RF) in order to determine the quantitative parameters characterizing the tissue. The QUS technique was also used to monitor the breast cancer response to chemotherapy. For example, Lin et al. used animal models of breast cancer to show that the spectral analysis of ultrasonic echoes provides a way to assess the tumor response to chemotherapy [12]. Similarly, Sannachi et al.

carried out a study on a group of 30 tumors and showed that the use of a combination of quantitative measures: average scatterer diameter (ASD), and average acoustic concentration (AAC) allows for the differentiation of responding and non-responding patients, with a sensitivity of 82% and a specificity of 100% after the fourth week of treatment [13]. In later studies, Sannachi et al., based on a combination of quantitative parameters: texture, and molecular features, predicted tumor response to NAC with an accuracy of 79, 86, and 83% at weeks 1, 4, and 8 of the therapy, respectively [14]. Sadeghi-Naini et al. presented clinical trials using QUS on a group of 100 patients [15]; the authors used the mean values of quantitative parameters (mid-band fit, slope, intercept, spacing among scatterers, ASD, and AAC) as well as textural features of parametric maps based on these parameters. Their results show that QUS imaging is able to predict the outcome of chemotherapy in patients with breast cancer with an AUC (area under the receiver-operating characteristic curve) of 0.80 and 0.89, respectively, 4 and 8 weeks after the start of the treatment.

This paper presents the results of monitoring changes occurring in breast tumors during neo-adjuvant chemotherapy in a group of 24 tumors. The QUS parameters: integrated backscatter (IBSC), and the shape parameter of homodyned K distribution (H-K) were used as markers of tissue changes occurring during NAC. In the case of responding tumors, NAC therapy reduces the number of cancer cells and causes significant changes in the structure of the remaining malignant cells and stroma tissue. The value of the H-K distribution shape parameter depends on the effective number of scatterers (ENS) in the resolution cell. We therefore hypothesize that changes associated with NAC therapy cause variations in the spatial density of scatterers and their scattering properties and thus affect the statistics of the tissue backscatter. This should affect the value of ENS. The H-K distribution has not previously been used to monitor the effects of NAC. The second quantitative parameter,

IBSC, determines the amount of scattered energy and depends on the number, size, and elastic properties of the scattering elements in the tissue.

Parallel to the studies on the usefulness of QUS parameters the effect of NAC therapy on tumors was evaluated based on tumor size assessment. The accuracy of tumor response assessment using the proposed quantitative and clinical methods was compared to the histopathological results of the post-operative study. An attempt was made to predict the effectiveness of monitoring NAC effects with QUS before the end of the treatment, i.e. after subsequent cycles of chemotherapy.

Based on the results, the statistical properties of ultrasonic scattering are important in monitoring the effects of chemotherapy and, together with the integrated backscatter, allow forecasting the final result of chemotherapy after the second dose of NAC, with a satisfactory result (Area under the ROC Curve = 0.82).

# Materials and methods

## Patients

The study was carried out in the Department of Radiology, Maria Skłodowska-Curie Memorial Institute of Oncology in Warsaw, Poland.

The study included patients with breast cancer qualified for neoadjuvant chemotherapy (NAC) prior to mastectomy. The group consisted of patients with operable (T2-3, N0-2, M0) or locally advanced (T4a-d, N0-2, M0) tumors.

The Institutional Review Board approved the study protocol. All procedures performed in the study that involved human participants were in accordance with the guidelines set by the 1964 WMA Declaration of Helsinki and its later amendments or comparable ethical standards. All patients signed the informed consent for breast US examination and for ‘backscatter US’ statistical studies.

The US examinations were performed in 16 patients aged 32 to 83 years (median 53.5). Three patients had bifocal lesions and two patients had three-focal lesions, resulting in a total of 24 tumors for examination. Data from each cancer were processed separately. One tumor was non-specific type (NST) with ductal carcinoma in situ (DCIC), while the remaining tumors were NST. Lesions were verified as invasive carcinoma NST G2 (15), G3 (6), and G1 (3). There were five luminal A cancers, five luminal B Her2+, 10 luminal B Her 2–, two TNBC, and two HER2+. The NAC treatments were administered according to the international guidelines in the protocol: AC (doxorubicin, cyclophosphamide), taxol and trastuzumab were used. In one patient with a history of breast cancer, AT (doxorubicin, docetaxel) was used. All patients underwent a simple mastectomy with lymphadenectomy.

## Ultrasound data acquisition

The B-mode ultrasound (US) images and raw Radio Frequency data (RF) from the breast tumors were acquired using an ultrasonic scanner (Ultrasonix Sonix Touch-Research, Ultrasonix Medical Corporation, Richmond, BC, Canada) with a linear array transducer L14-5/38 and a transmit frequency of 7 MHz (-6 dB bandwidth range of 4-9 MHz). The focus was set to the lower part of the tumors. Data samples were recorded with a sampling frequency of 40 MHz and a 16-bit precision. Each data frame consisted of 510 lines, which corresponded to ~40 mm. The measurement scheme included ultrasound data registration in four tumor planes (radial, radial + 45°, anti-radial, anti-radial +45°). The ultrasound examinations were performed in accordance with the American College of Radiology BI-RADS guidelines [16].

The region of interest (ROI) contoured around the tumor was determined by the same, experienced radiologist during each scan. The size of the tumor in the longest dimension was also measured.

The first scan was acquired before chemotherapy and was used as the baseline data. Subsequent ultrasound examinations were performed 1 week after each course of NAC; the patient was monitored for 5 to 6 months. All patients underwent the first four cycles of NAC, but some patients did not take part in subsequent stages of chemotherapy due to the oncologist's decision to perform a mastectomy or in the event of complete disappearance of the tumor.

## Quantitative ultrasound parameters

Two QUS parameters were used, the integrated backscattering coefficient (IBSC) and the shape parameter of the homodyned K distribution. The hypothesis underlying their use is that changes in the structure and function of the tumor associated with the effects of chemotherapy result in changes in the basic physical properties of the tissue. These properties can be quantified with indicators dependent on the frequency and the backscatter statistics. The IBSC quantifies the energy of backscattering and depends on the size and physical properties of the scatterers, their concentration, and the randomness of parameters. The shape parameter of the homodyned K distribution depends on the effective number of scattering (ENS) elements in the resolution cell.

The distribution of IBSC and ENS values in the tumor was presented in the form of parametric maps, which were generated using the sliding window technique. The window was moving pixel by pixel, in the horizontal and vertical directions, in the area of the ROI covering the entire tumor area. The parameters IBSC and ENS were calculated based on ultrasound RF data from each window. The window had dimensions of 3 mm by 3 mm, which meets the window size requirements for obtaining reliable scatterer property estimates [17,18]. The tumor response to the treatment was analyzed by assessing the mean IBSC values and the mean ENS values obtained in all windows of the four parametric maps corresponding to the four tumor sections.

The RF signal analysis and the estimation of the quantitative ultrasonic parameters were performed off-line using in-house software written in the Matlab environment (Mathworks, Natick, MA, USA).

## Integrated backscatter coefficient

The integrated backscatter coefficient was derived from estimates of the backscatter coefficient (BSC), which is defined as the differential scattering cross section per unit solid angle at 180° per unit volume. The BSC was determined using the reference phantom method proposed by Yao et al. [19]. The technique allows limiting the impacts of system-dependent effects such as the system transfer function or diffraction artifacts by using data from the well-defined reference phantom where the attenuation and the backscatter coefficient are known. In the present study, the tissue mimicking phantom (1126 B, Dansk Phantom Service, Denmark) was used as reference, with an attenuation coefficient of 0.5 [dB·MHz$^{-1}$·cm$^{-1}$] and a speed of sound of 1540 [m/s]. The BSC of the reference phantom ($\sigma_R(f)$) was measured using the methods described by Ueda and Ozawa [20]. The backscatter coefficient ($\sigma(f)$), using the reference phantom technique, is obtained by the following formula:

$$\sigma(f) = \sigma_R(f)\frac{S_t(f)}{S_R(f)}e^{4\alpha(f)d},$$

where $f$ is a frequency, $S_t(f)$and $S_R(f)$ are the averaged power spectra of signals from the tissue and the reference phantom, respectively, $\alpha(f)$ is the attenuation coefficient expressed in Nepers, and $d$ is the distance traveled in the phantom to the considered window position. The attenuation coefficient of breast tissue was assumed to be 1 [dB·MHz$^{-1}$·cm$^{-1}$] [21]. The integrated backscatter coefficient (IBSC) was then calculated by integrating the backscatter coefficient within the transmitter bandwidth (4 MHz - 9 MHz) [22].

## Envelope statistics

The envelope statistics was based on the homodyned K distribution, which was first introduced by Jakeman [23]; this model of backscatter was chosen because of its flexibility. The high usability of homodyned K distribution stems from the ability to describe statistics of envelopes of signals under varying conditions. This means that homodyned K distribution may be used when the number of scatterers is large and when they are uniformly distributed as well as when the number of scatterers is low or they are organized in periodical structures or in case of coherent component sources. Additionally, the shape parameter of the homodyned K distribution has shown that it may be used as an ultrasonic biomarker which characterizes the tissue microstructure [24]. The shape parameter ENS, also called “effective number of scatterers", is defined as ENS= $N(1-\vartheta)$, where $N$ is the real number of scatterers and $\vartheta$ describes the level of clustering. The method proposed by Hruska and Oelze was used to determine ENS [25].

## Statistical analysis

For statistical analysis, QUS parameters and the size parameter were expressed as percent changes in their values in relation to the initial values before treatment. The relative values of the IBSC and ENS for each patient and each treatment stage were the features of the classification of the therapy effectiveness. Three QUS-based classifiers were used: two using the individual QUS features and one (referred to as IBSC + ENS) based on their linear combination obtained with the use of the Linear Discriminant Analysis (LDA) [26]. Additionally, a tumor size-based classifier was included in the analysis.

The statistical significance of the classifiers was assessed using *p*-values obtained with a two-sided Wilcoxon rank sum test. The classifiers were cross-validated through the leave-one-out technique [27]. Evaluation of the results was based on the area under the receiver operating characteristic (ROC) curve (AUC) [28]; these values were determined for data acquired after each chemotherapy course. The results were analyzed for changes in classification effectiveness as a function of the therapy progress. All calculations were done using Matlab® 2018a (The MathWorks, Inc., Natick, MA).

# Results and discussion

## Pathologic responses to treatment

Tumor tissue specimens were histologically evaluated for assessment of the effects of NAC. The experienced histopathologist (20 years in an Oncology Institute focusing on breast cancer) estimated the percentage of residual malignant cells (RMC) from the localized and removed tumors. Miller-Payne scoring [29] was the basis for qualifying the examined tumors in the pathological responding (PR) or non-responding (N-PR) group. Tumors with total cell loss of up to 30% (Miller-Payne G1 or G2 score) were assigned to N-PR, while the remaining tumors (Miller-Payne G3 - G5 score) were assigned to the PR group. In the group of tumors studied, there were 5/24 cases with RMC equal to 100%, classified as non-responding (N-PR). The remaining cases (19/24) were assigned to the group responding pathologically (PR), including 6/19 cases of complete response with an RMC of 0%.

Microscopic analysis of the samples shows the changes occurring as a result of the NAC, which is important for the scattering of ultrasounds. In the case of the PR tumors, we observed a decrease in cellularity, formation of fibrosis, and stromal edema. Microscopic images of cancer samples for responding and non-responding tumors are shown in Fig 1. The RMC values for these tumors after NAC treatment were 0 and 100%, respectively. Microscopic evaluation of the specimens revealed the presence of stroma-filled tissue (pink staining) with small isolated patches of glands (purple staining), demonstrating therapeutic effects (Fig 1a). In Fig 1b, high cellularity with low stromal collagen density is visible. Assessment of the usefulness of QUS parameters and of the tumor size to predict the effects of chemotherapy was based on a comparison with RMC results.

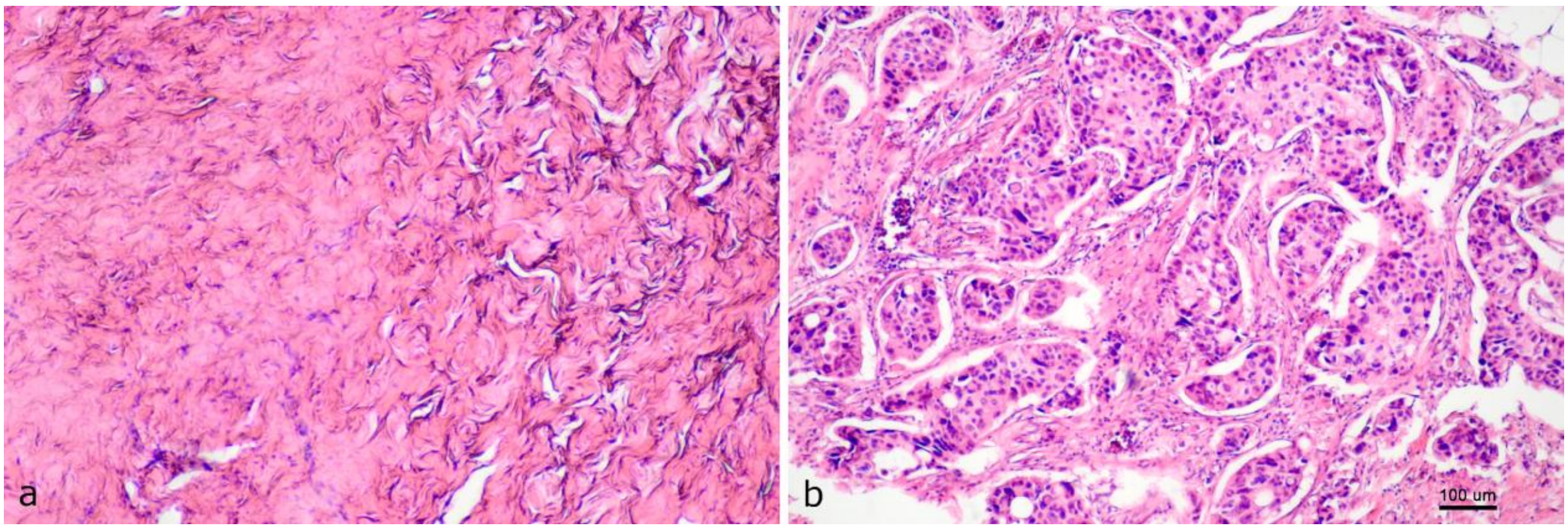

**Fig 1. Histopathological images of the responding (PR) and non-responding (N-PR) breast tumors**. In a PR tumor, there was a lack of tumor cells (a), in the N-PR tumors, malignant cells persisted (b).

## Tumor size response

Changes in the longest dimension of the tumor during subsequent chemotherapy courses are shown in Figure 2 in relation to the tumor length before therapy. For patients who had completed the NAC earlier, the results from the data collected after the last chemotherapy were presented along with the results of other cancers collected at later stages. This also applies to results for ENS and IBSC parameters.

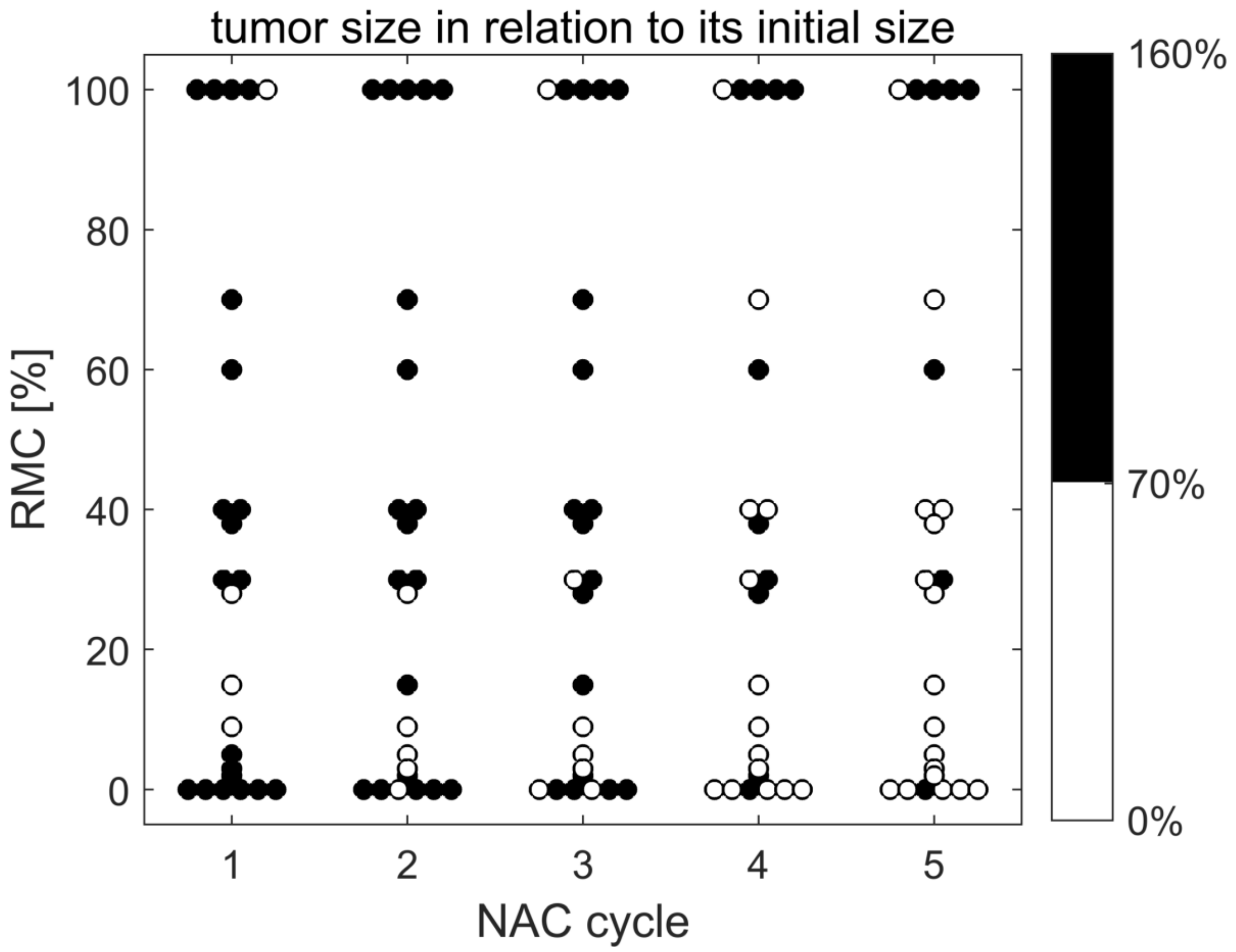

**Fig 2. Relative changes in maximum tumor length depending on the postoperative RMC value and the NAC dose number**. Tumors whose maximum size has dropped below 70% of the pre-treatment dimension are marked white; other tumors in black.

The tumors were ranked according to the RMC value, which is marked on the vertical axis. On the horizontal axis, subsequent courses of chemotherapy were marked; the 24 circles correspond to 24 tumors. Those whose length has decreased by 30% or more are marked in white, while the tumors marked in black represent smaller changes or tumor growth. The 30% decrease was adopted as a limit value according to recommendations of the RECIST 1.1 (Response Evaluation Criteria in Solid Tumors [30]), dividing tumors into responders and non-responders depending on their maximum length change. Although ultrasonography is not recommended in the RECIST 1.1 guideline as a tool for assessing tumor changes during chemotherapy, these criteria are used in other scientific work [31].

In the group of tumors responding to NAC, according to histopathological assessment, there was a tendency to reduce the maximum tumor size along with subsequent doses of chemotherapy. After the end of the therapy, only in 3 out of 19 cases there were no changes to qualify the cancer, based on size reduction, as NAC-responsive tumors. In the group of N-PR tumors in one case, the assessment based on changes in size was inconsistent with an RMC-based assessment. The lack of size-based responses, when the response was positive, results from the fact that the invasion of fibrous stroma, caused by chemotherapy, results in the overestimation of tumor residue in the ultrasound [32].

## QUS parametric maps

Cycles of representative B-mode ultrasound images for responding (0% RMC) and non-responding (100% RMC) tumors are shown in Figure 3, together with the parametric overlays of the IBSC and ENS maps. They were calculated based on ultrasound data collected immediately before NAC treatment and 1 week after each NAC course. The pixel color reflects the parameter value and the position depends on the location of the sliding window. Analysis of the IBSC map set revealed a clear upward trend in the IBSC values of responding tumors (Fig 3, first column). In the case of non-responding tumors, no such changes were observed (second column). For the responding tumor, the ENS values decreased as the treatment progressed (Fig 3, third column). In the case of a non-responding tumor, there was no clear trend in ENS changes (last column).

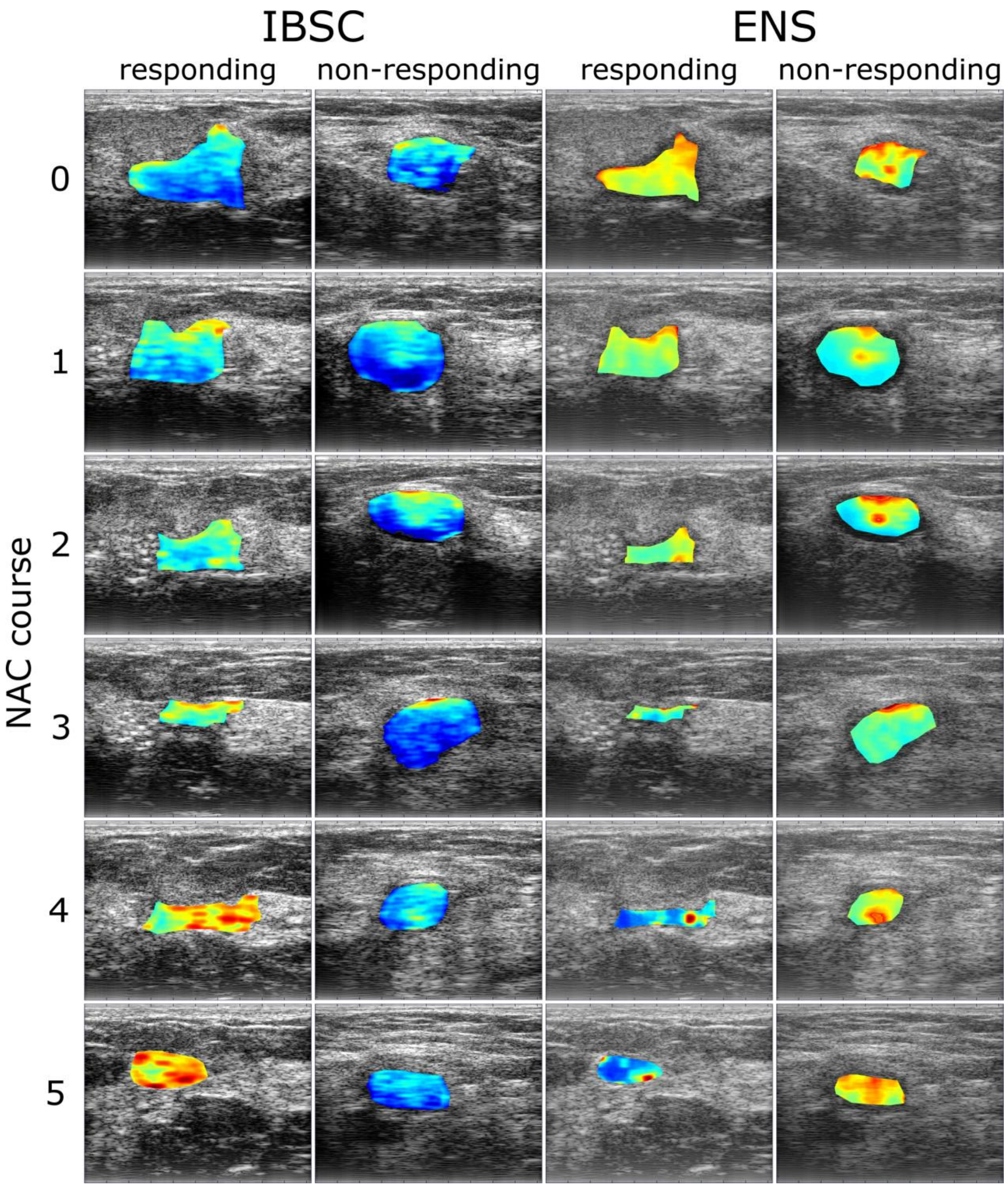

**Fig 3. Examples of ultrasound B-mode images with IBSC and ENS parametric overlays**. Ultrasound data were obtained from a responding tumor (0% RMC) and non-responding tumor (100% RMC) prior to (first row) and 1 week after every NAC course (successive rows). The blue color indicates low values of both quantitative parameters, IBSC or ENS, while the red color indicates high values.

## Responses of QUS parameters

Quantitative ultrasound parameters, mean IBSC and ENS values, were determined from the parametric maps. Changes in quantitative parameters during subsequent chemotherapy courses are shown in Figure 4 in relation to the parameter values before therapy. The form of the presentation is analogous to the presentation of the results of changes in maximum tumor size with one difference. Relative changes of parameters were divided into three ranges (presented in color-bar in Fig 4), reflecting the extent of the increase or decrease of the parameter under consideration.

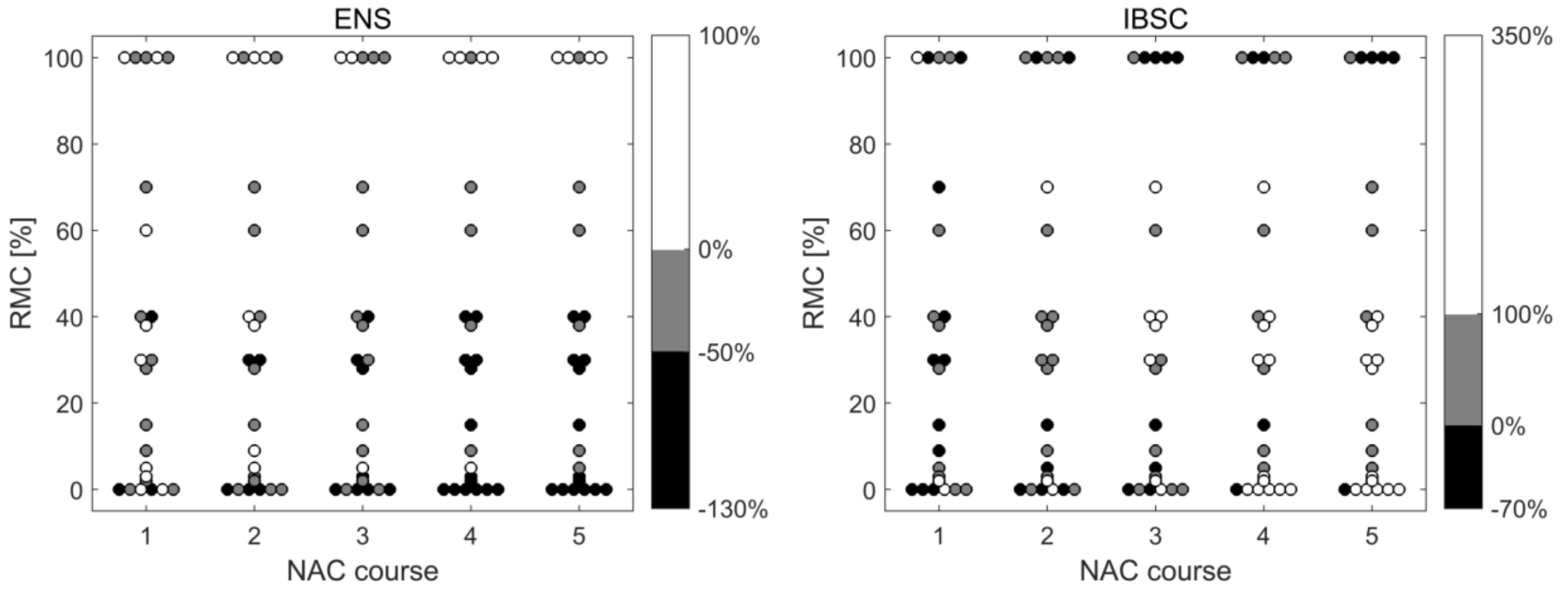

**Fig 4. Relative changes in the values of ENS and IBSC, depending on the value of postoperative RMC and the number of NAC doses.**

The ENS coefficient decreased with successive NAC doses for low-RMC tumors, i.e. for responding tumors. In the case of non-reacting tumors, ENS increased or did not change. This ENS behavior implies a changing number of scatterers in the resolution cell. The comparison of the histopathological findings of tumors that responded to the treatment before and after NAC suggests that other scattering structures are the main sources of scattering after chemistry doses. Changes occurring in cancer tissue during chemotherapy concern both changes at the cell level and changes in the stroma tissue structure. The nuclear structure of the cell and its physical parameters such as density, elasticity, and viscosity, are changing [33], and the condensation of the cell's nucleus during apoptosis increases the ultrasound scattering [34]. On the other hand, the stromal microenvironment undergoes changes in the blood vessel architecture and the extracellular matrix composition [35,36]. In the stroma, fibrosis, collagenization, and microcalcification are associated with structures strongly scattering ultrasounds. It can be assumed that before therapy, ultrasound scattering on tumor cell clusters plays a significant role in the scattered signal. In the case of effective chemotherapy, the newly formed structures of the repair processes, such as excess fibrous connective tissues, significantly contribute to the scattering of the ultrasound; based on their size and mechanical properties, they are scatterers influencing the value of the ENS parameter.

In one case, in the group of tumors that did not respond, a reduced ENS value after NAC treatment was found. It was an invasive NST breast cancer with in situ ductal carcinoma (DCIS), while the remaining tumors were NST type. In DCIS tumors, cancer cell proliferation occurs inside the ducts and does not infiltrate the surrounding tissue. Thus, it can be assumed that in this case a small decrease in the EDS parameter was due to the fact that changes in tumor cells during NAC occurred in the ducts, while the surrounding tissues and stroma did not show similar changes as in infiltrated cancer.

The increase in IBSC value was the reaction of cancerous tissue to NAC in the case of a positive histopathological result. In the group of non-responders, IBSC decreased or remained unchanged. After completion of the NAC treatment, only in one case of the non-responding tumors, a small increase in the IBSC value was found, and in one case of responding tumors, this value decreased. This behavior of IBSC confirms the results of another study which showed an increase in the backscattering coefficient of ultrasound in the breast tumor, in the 4-9 MHz range, as a positive response to chemotherapy [13].

The QUS parameters are also shown in the IBSC-ENS plane (Fig 5), with the division into responding and non-responding tumors classified according to histopathological results. With the continuation of therapy, the group of respondents (white markers) clearly shifts to the right, to high values of IBSC growth and down to the values indicating the reduction of ENS, while cases of cancer in the non-responding group (black markers) remain almost stationary. The increase in IBSC and the decrease in ENS of the responding tumor shown on the parametric maps in Fig 3 are therefore confirmed in a larger sample. Similarly, no change in the QUS parameters of the non-responding tumor is representative for non-responding cases.

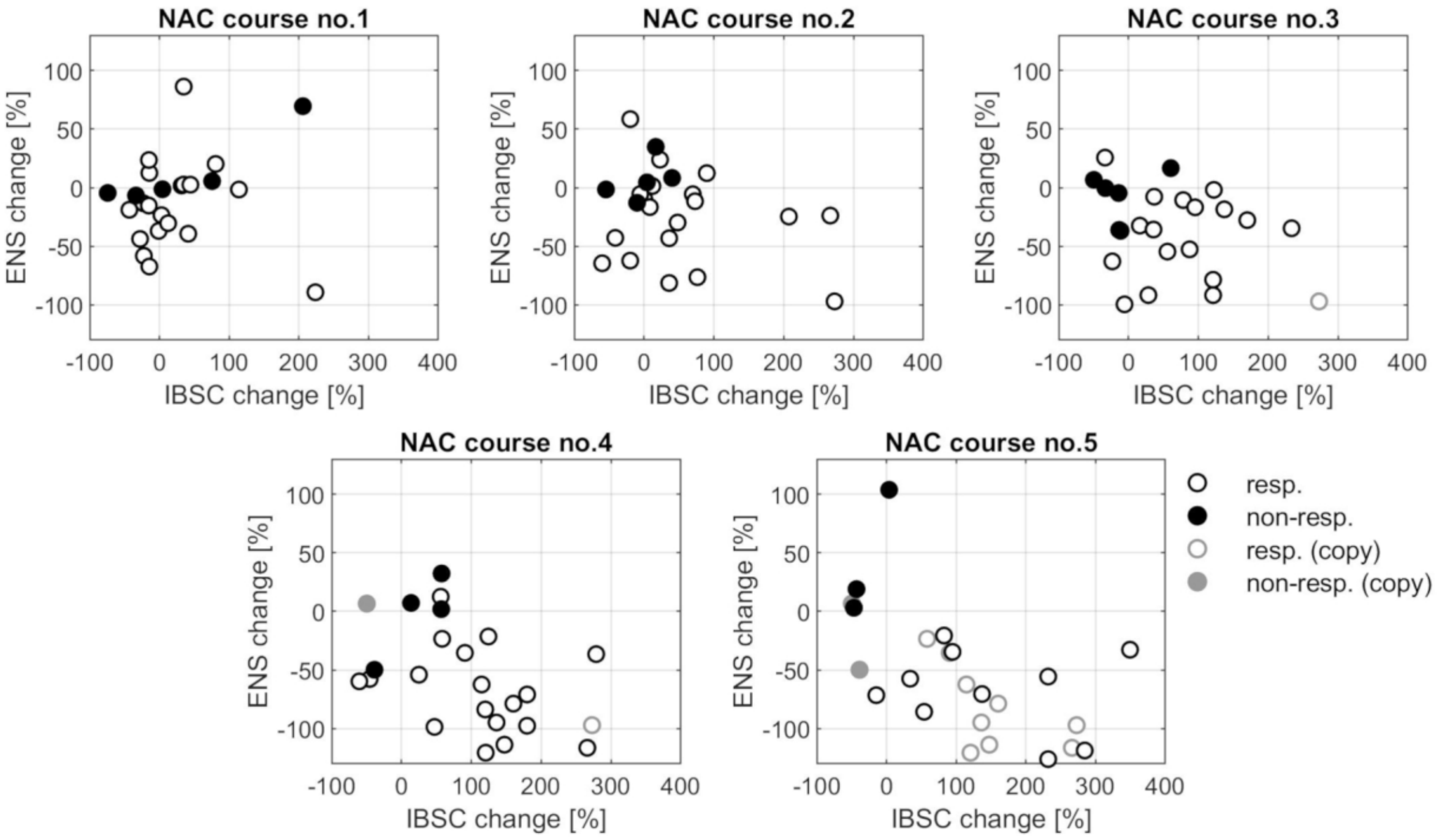

**Fig 5. Scatter plots of pairs of features, ENS vs. IBSC, for all PR and N-PR tumors for each treatment stage**. Some patients have undergone less than five NAC courses; in these cases, the results obtained after their last NAC course were copied into the further results (less intense markers).

## Classification and statistical analysis results

Statistical analysis was performed to distinguish between response cases and no response based on ENS, IBSC, IBSC + ENS, or tumor size, features used as biomarkers for NAC response. The statistical significance of the differences between the NAC-responsive groups and the groups that did not respond was evaluated with the *p*-values shown in Table 1, where statistically significant differences ($p < 0.05$) are marked in gray.

**Table 1 *p*-values for treatment response classification based on ENS, IBSC, IBSC + ENS, and tumor size at each NAC treatment stage. *p* < 0.05 is marked in gray.**

| NAC course # | IBSC | ENS | IBSC+ENS | Size |
|---|---|---|---|---|
| 1 | 0.94 | 0.20 | 0.48 | 0.52 |
| 2 | 0.20 | 0.05 | 0.03 | 0.78 |
| 3 | 0.03 | 0.04 | 0.01 | 0.85 |
| 4 | 0.08 | 0.04 | 0.03 | 0.14 |
| 5 | 0.01 | 0.01 | 0.01 | 0.05 |

The progressive separation of responding and non-responding groups (Fig. 5) raises the question about the treatment step which allows an effective assessment of tumor response. The QUS parameters, individually as well as their linear combination based on the LDA analysis, were used as classifiers of tumor response to NAC. The classification based on the largest tumor size was also assessed. The obtained AUC values calculated at each NAC stage are shown in Fig 6 and Table 2.

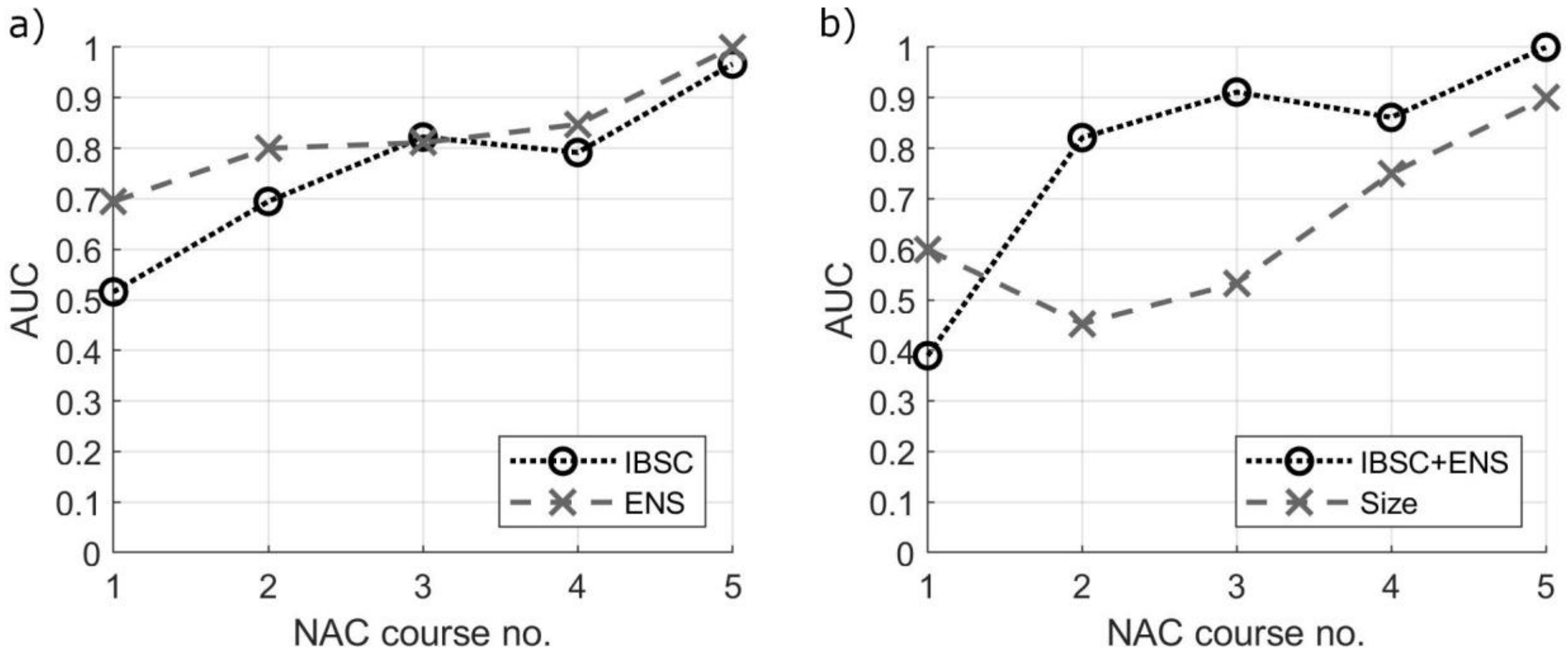

**Fig 6. Area under the ROC curve (AUC) values for treatment response classification at each treatment stage.** a) individual QUS parameters, IBSC and ENS and b) linear combination of IBSC and ENS (IBSC+ENS) and the largest tumor size.

**Table 2 AUC values for treatment response classification based on ENS, IBSC, IBSC + ENS, and tumor size at each NAC treatment stage.**

| NAC course # | IBSC | ENS | IBSC+ENS | Size |
|---|---|---|---|---|
| 1 | 0.52 | 0.69 | 0.39 | 0.60 |
| 2 | 0.69 | 0.80 | 0.82 | 0.45 |
| 3 | 0.82 | 0.81 | 0.91 | 0.53 |
| 4 | 0.79 | 0.85 | 0.86 | 0.75 |
| 5 | 0.97 | 1.00 | 1.00 | 0.90 |

Each classifier started with poor results at the first NAC course (p-value > 0.05 and AUC ≈ 0.5, except ENS, where AUC ≈ 0.7), along with additional NAC doses, improved, and ended up as highly effective after the fifth dose of NAC (p-values < 0.05 and AUC ≈ 1). The ENS performed better than the IBSC (see Fig 6a). A combination of QUS parameters allowed improving the AUC after the second, the third, and the fourth NAC course (Fig. 6b).

The main difference in the efficiencies of classification based on the IBSC + ENS composite classifier and the classification based on the size of tumors is the different stage of NAC in obtaining a similar AUC (Fig 6b). This difference, however, is extremely important because the early diagnosis of N-PR tumors is fundamental to personalize the treatment and allows oncologists to make an early decision depending on the effectiveness of the NAC.

For an assessment based on tumor size, often used in clinical evaluation, effective classification is only possible after NAC stage 4 and 5 (AUC = 0.75 and 0.90 respectively). For a combined QUS classifier a comparable AUC is available after the second and third NAC courses respectively. This delay in tumor size reduction results from the lower rate of the tissue repair process, manifested by the disappearance of the tumor, compared to the changes taking place in the cells and tissue structure.

# Conclusions

The results obtained suggest that ENS provides useful information to monitor NAC. Thus, the statistical properties of ultrasonic backscattering were important in monitoring chemotherapy effects on an equal footing with IBSC. Our studies have shown that tumor quantitative parameters when maintained at the same level after NAC courses predict a poor response to the treatment. At the same time, the decrease in ENS and the increase in IBSC are characteristic of a positive tumor response. The presented results suggest that quantitative ultrasound information can characterize the tumor's pathological response better and at an earlier stage of therapy than the assessment of the reduction of its dimensions based on ultrasound imaging. Statistical analysis proved that, after the second chemotherapy the final result can be effectively predicted based on linear combination of changes in the values of the ENS and IBSC. This significant result, if confirmed in a larger group of cases, suggests that the introduction of ultrasound backscatter statistics parameters to monitor the effects of chemotherapy may increase the effectiveness of monitoring and contribute to a better personalization of the NAC therapy.

# Supporting information

**S1 Table. Relative changes in IBSC and ENS parameters, and tumor size of all 24 cancers undergoing subsequent NAC courses.**

| Tumor no. | Response type | IBSC relative change [%] | | | | | ENS relative change [%] | | | | | Size relative change [%] | | | | |
|---|---|---|---|---|---|---|---|---|---|---|---|---|---|---|---|---|
| | | course no. 1 | course no. 2 | course no. 3 | course no. 4 | course no. 5 | course no. 1 | course no. 2 | course no. 3 | course no. 4 | course no. 5 | course no. 1 | course no. 2 | course no. 3 | course no. 4 | course no. 5 |
| 1 | non-resp. | 206 | 17 | 60 | 57 | 4 | 70 | 35 | 17 | 32 | 104 | 35 | -15 | -40 | -35 | -35 |
| 2 | non-resp. | -34 | -10 | -50 | - | - | -7 | -13 | 7 | - | - | 8 | -8 | 8 | - | - |
| 3 | non-resp. | 4 | 40 | -13 | -39 | - | -1 | 8 | -36 | -50 | - | -4 | -22 | -26 | -30 | - |
| 4 | non-resp. | 76 | 4 | -14 | 57 | -47 | 6 | 5 | -5 | 2 | 3 | 20 | -20 | -7 | 0 | 13 |
| 5 | non-resp. | -74 | -55 | -33 | 14 | -43 | -5 | -2 | 0 | 7 | 19 | -38 | -19 | -19 | -23 | -19 |
| 6 | resp. | -22 | -20 | -23 | -46 | -15 | -58 | -62 | -63 | -58 | -71 | 0 | -10 | -30 | -40 | -60 |
| 7 | resp. | 41 | 12 | 122 | 25 | 34 | -39 | 1 | -2 | -54 | -57 | 59 | 0 | -4 | -52 | -70 |
| 8 | resp. | -15 | 36 | 121 | 180 | 232 | -67 | -43 | -79 | -71 | -56 | 7 | 2 | -24 | -59 | -63 |
| 9 | resp. | -28 | -41 | -11 | -60 | 54 | -44 | -43 | -37 | -60 | -86 | -41 | -26 | -22 | -41 | -44 |
| 10 | resp. | -43 | 267 | 170 | 124 | 95 | -19 | -24 | -28 | -21 | -34 | 32 | 12 | -12 | -36 | -52 |
| 11 | resp. | -1 | 23 | 16 | 91 | - | -37 | 24 | -32 | -35 | - | -32 | -45 | -58 | -55 | - |
| 12 | resp. | -23 | 0 | 78 | 136 | - | -13 | -10 | -11 | -95 | - | 4 | -13 | -8 | -38 | - |
| 13 | resp. | 3 | 8 | 28 | 47 | 284 | -23 | -17 | -92 | -99 | -119 | -41 | -41 | -19 | -19 | -41 |
| 14 | resp. | -15 | 36 | 121 | 180 | 232 | 13 | -81 | -92 | -98 | -126 | -28 | -28 | -38 | -41 | -45 |
| 15 | resp. | -15 | 76 | 36 | 121 | - | -15 | -76 | -36 | -121 | - | -17 | 0 | 17 | 0 | - |
| 16 | resp. | 81 | -20 | -34 | 56 | 82 | 20 | 58 | 26 | 13 | -21 | -14 | -33 | -43 | -48 | -62 |
| 17 | resp. | -15 | -60 | -6 | 120 | 137 | 24 | -65 | -100 | -84 | -70 | -10 | -30 | -20 | -20 | -20 |
| 18 | resp. | 224 | 273 | - | - | - | -89 | -97 | - | - | - | 20 | -20 | - | - | - |
| 19 | resp. | 35 | 90 | 137 | 279 | 349 | 86 | 12 | -18 | -36 | -33 | -9 | 0 | -14 | -27 | -50 |
| 20 | resp. | 32 | -5 | 37 | 266 | - | 2 | -5 | -8 | -116 | - | 23 | -15 | -23 | -46 | - |
| 21 | resp. | 12 | 48 | 87 | 115 | - | -30 | -30 | -53 | -62 | - | 17 | 17 | -25 | -42 | - |
| 22 | resp. | 35 | 70 | 95 | 58 | - | 3 | -6 | -17 | -23 | - | 6 | 3 | 0 | -6 | - |
| 23 | resp. | 44 | 73 | 56 | 160 | - | 3 | -12 | -55 | -79 | - | -20 | -35 | -35 | -40 | - |
| 24 | resp. | 114 | 207 | 233 | 147 | - | -2 | -25 | -35 | -114 | - | 0 | -9 | -18 | -18 | - |